\documentclass[pdftex,twocolumn,epjc3]{svjour3}          

\RequirePackage[T1]{fontenc}

\smartqed  

\RequirePackage{graphicx}
\RequirePackage{mathptmx}      
\RequirePackage{flushend}
\RequirePackage[numbers,sort&compress]{natbib}
\RequirePackage[colorlinks,citecolor=blue,urlcolor=blue,linkcolor=blue]{hyperref}

\journalname{Eur. Phys. J. C}

\begin{document}

\title{On the hypotheses of Penrose's singularity theorem under disformal transformations}


\author{Eduardo Bittencourt\thanksref{e1,addr1}
        \and
        Gabriel G. Carvalho\thanksref{e2,addr2}
        \and
        Iarley P. Lobo\thanksref{e3,addr3,addr4}
        \and
        Leandro Santana\thanksref{e4,addr1}
}

\thankstext{e1}{e-mail: bittencourt@unifei.edu.br}
\thankstext{e2}{e-mail: ggc5@cin.ufpe.br}
\thankstext{e3}{e-mail: iarley\_lobo@fisica.ufpb.br}
\thankstext{e4}{e-mail: leandro09@unifei.edu.br}

\institute{Federal University of Itajub\'a, Itajub\'a, Minas Gerais 37500-903, Brazil\label{addr1}
          \and
          Centro de Inform\'atica, Federal University of Pernambuco, Recife, Pernambuco,  50740-560, Brazil\label{addr2}
          \and
          Federal University of Para\'iba, Jo\~ao Pessoa, Para\'iba, 58059-900, Brazil\label{addr3}
         \and
          \emph{Present Address:} Federal University of Lavras, Lavras, Minas Gerais 37200-000, Brazil\label{addr4}
}

\date{Received: date / Accepted: date}

\maketitle

\begin{abstract}
We analyze how the hypotheses of Penrose's singularity theorem (1965) are modified by the action of disformal transformations (defined in terms of light-like vectors) upon a given space-time metric. In particular, we investigate the transformation of the null energy condition and the existence of closed trapped surfaces in such scenario, in order to derive conditions upon the background metric and the disformal vector that guarantee the validity of Penrose's theorem for disformal metrics. Then, we explain how to apply this technique for static and spherically symmetric space-times in general.
\end{abstract}

\section{Introduction}
Black holes and big bangs are examples of singularities that inspire curiosity even in the realm of popular science. Commonly, they are depicted as ``catastrophic'' events such that time and space behave in a counter-intuitive manner near their boundaries: the matter content increases enormously the gravitational field and the curvature diverges. Indeed, they represent what is called  a space-time singularity. In the Sixties, there was a deep discussion about the limits of predictability in general relativity due to the appearance of such singularities. Initially, physicists analyzed the vicinities of a space-time singularity by searching for generic analytic solutions (see \cite{belinski} and references therein) wondering whether singularities would appear only in very special cases with high degree of symmetry. But, soon after, Penrose and Hawking published a few papers demonstrating that under certain circumstances space-time singularities are unavoidable \cite{haw_ellis}. Their singularity theorems are based upon the geodesic incompleteness of a given space-time, indicating that a given manifold has a singularity if there exists at least one incomplete geodesic curve parameterized by an affine parameter. Usually, such theorems are interpreted as an indicative of the invalidity of general relativity near singularities, suggesting the need for a more sophisticated theory of gravitation.

In this paper, we study how the hypotheses of the singularity theorems can be modified by some transformation acting on the space-time metric. In fact, we shall analyze the original Penrose's theorem (1965) which will be used along the text \cite{haw_ellis,penrose}: ``Space-time cannot be null geodesically complete if: {\bf (i)} $R_{\mu\nu}k^{\mu}k^{\nu}\geq0$ for all light-like vector $k^{\mu}$; {\bf (ii)} there is a non-compact Cauchy surface; {\bf (iii)} there is a closed trapped surface''. We should emphasize that the singularity theorems obey a general structure described in Ref.\ \cite{Senovilla:2018aav}, in which the hypotheses concern: {\bf (i)} an energy condition; {\bf (ii)} a causality condition; and {\bf (iii)} a boundary or initial condition. All these ingredients are present in any formulation of the singularity theorems. Notwithstanding, the way causality conditions are modified when we perform a transformation on the space-time metric cannot be predicted unless we know the topology change caused by the transformation a priori. Of course, any singularity in the elements involved by the mapping itself will play a crucial role in the determination of the global aspects of the resulting space-time. However, it would require a sort of a topological classification of the space-times, which is out of our scope. Furthermore, there are well-known theorems in topology demonstrating that a complete classification of manifolds for $\dim {\mathcal M}\geq4$ is not possible. For the sake of comparison, there are recent papers looking for it in special classes of space-times (for instance, see \cite{martin} and references therein). Therefore, we shall deal with a restricted class of metrics that preserves the condition {\bf (ii)},  which is a fairly reasonable assumption from the physical point of view. As we shall see, the tools we develop here can be naturally used to investigate the energy and boundary conditions in the context of Penrose's 1965 singularity theorem \cite{penrose} and a similar examination can be done for other cases with more or less restrictive conditions on the space-time.

In this vein, we shall investigate the appearance of singularities in a given space-time when we consider another space-time related to the former via a disformal transformation. A space-time is a pair $(\mathcal{M},g)$, i.e a smooth manifold and a metric with Lorentzian signature. Then, by definition, $(\mathcal{M}, \hat{g})$ is a different space-time. The relation between $g$ and $\hat{g}$ will determine the events (points in $\mathcal{M}$) that may still be allowed to happen should such a transformation occur. Considering $\mathcal{M}$ with two metrics is acceptable from the mathematical viewpoint, since a metric is just a tensor field satisfying an appropriate definition. Nonetheless, from the physical standpoint, it is worth to consider them as being two different space-times. Thus, we shall use this kind of transformation to scrutinize the hypothesis of the singularity theorem quoted above in order to identify the precise conditions that could introduce (or remove) a singularity on the space-time when the transformation is performed. In practice, this analysis can be applied to understand the fact that there are different solutions of Einstein's field equations for the same matter content and related via a disformal transformation where some of them are singularity-free and the others are singular.
\par
The motivation for dealing with disformal transformations lays on the increasing literature on the subject as an alternative to solve current problems of gravity within the framework of general relativity. For instance, at very high energy, near the Planck energy, the disformal transformations have been used to introduce an energy dependent space-time metric as it is the case of Rainbow Gravity \cite{magueijo04,clb16} and doubly special relativity \cite{amelino01}, in order to seek for phenomenological effects of quantum gravity. Aiming at explaining the dark matter and dark energy issues, the disformal transformations appear in modified Newtonian dynamics \cite{beken_mond}, scalar-tensor theories \cite{scalartheory,mota,ip,sak1,sak2}, Mimetic gravity \cite{rua,matarrese,sunny1,sunny2} and Horndeski theory \cite{miguel1,vernizzi,dario}. In field theory, there are also several applications of the disformal transformations \cite{uzan,yuan,brax1,brax2,nov_bit_gordon,nov_bit_drag}, with special attention to the framework introduced to describe geometrically some phenomena in particle physics \cite{bitt_nov_faci,nov_bit,erico12,erico13,bitt15}.
\par
This paper is summarized as follows. In Sec.\ II, we introduce the basic tools concerning disformal transformation that shall be used along the text. In Sec.\ III, we address few comments about the hypotheses of the singularity theorems and the conformal transformation, once it is a special case of the disformal one. Then, in Sec.\ IV, we study how disformal transformations change the energy condition and lead to the formation of closed trapped surfaces, showing that it is possible to decide whether a disformal metric $\hat g$ has a singularity using conditions strictly defined on the background metric $g$. Finally, in Sec.\ V, we apply our results to the particular case of static and spherically symmetric space-times. Along the text, the conventions will broadly follow Ref.\ \cite{poisson}. That is, the speed of light is set to unit, we use the Lorentz signature $(- + + +)$, and Einstein's summation conventions.


\section{disformal transformation definition}
Disformal transformations can be regarded as a generalization of conformal transformations. As such, they do not represent a change in coordinates, but a local change in the geometry instead. One might think of a conformal transformation as a smooth, {\it isotropic} and infinitesimal stretch at a point, whereas a disformal transformation is a smooth, {\it anisotropic} and infinitesimal stretch at a point. Given a space-time $(\mathcal{M},g)$, a light-like vector $V \in \Gamma(T\mathcal{M})$, where $\Gamma(T\mathcal{M})$ is the set of vector fields tangent to $\mathcal{M}$, and two space-time-dependent scalars $\alpha$ and $\beta$ with $\alpha>0$, we define a light-like disformal transformation $(\mathcal{M},g,V,\alpha,\beta)\longmapsto (\mathcal{M},\hat{g})$ as a change in geometry when the metric tensor changes according to
\begin{eqnarray}
\label{disf1}
\hat{g}(\ast,\cdot) = \alpha g(\ast,\cdot) + \beta g(V,\ast)\otimes g(V,\cdot).
\end{eqnarray}
The inverse of $\hat{g}$, namely $\hat{h}$, is given by
\begin{eqnarray}
\label{disf2}
\hat{h}(\ast,\cdot) = \frac{1}{\alpha}h(\ast,\cdot) -\frac{\beta}{\alpha^{2}} V(\ast)\otimes V(\cdot),
\end{eqnarray}
where $h=g^{-1}$ and $V$ is regarded as being a linear map from $\Gamma(T^{\ast}\mathcal{M})$ to $C^{\infty}(\mathcal{M})$, where $C^{\infty}(\mathcal{M})$ is the set of smooth functions defined on $\mathcal{M}$, and $\ast$ and $\cdot$ represent the placements of arbitrary vector fields on which the tensor fields involved must act. Since we are now dealing with a manifold endowed with two metric tensors, it is important to distinguish which metric tensor is being used when raising and lowering indices. One shall deal with this problem by explicitly writing the metric in all formulae in which indices are raised or lowered. It is easy to show that requiring $\alpha>0$ is enough to keep the Lorentzian signature of the disformal metric. Note that this kind of map between space-times can also be seen as a conformal transformation of the Kerr-Schild metrics \cite{stephani}, which is an old and well-understood topic in the context of general relativity.

In terms of a frame assigned by a local observer $\{x^{\mu}\}$, Eqs.\ (\ref{disf1}) and (\ref{disf2}) are written as
\begin{eqnarray}
\widehat{g}_{\mu\nu}&=&\alpha g_{\mu\nu}+\beta V_{\mu}V_{\nu},\label{cov}\\[2ex]
\widehat{g}^{\mu\nu}&=&\frac{1}{\alpha}g^{\mu\nu}-\frac{\beta}{\alpha^2}V^{\mu}V^{\nu}.\label{contrav}
\end{eqnarray}
It is straightforward to verify that the relation $\widehat{g}_{\mu\nu}\widehat{g}^{\nu\sigma}=\delta_{\mu}^{\sigma}$ holds true, where $\delta_{\mu}^{\sigma}$ is the Kronecker delta.

The key differences between conformal and light-like disformal transformations were studied by some of us in Ref.\ \cite{lc19}, where the reader will find the transformation formulae of some geometric quantities. For the sake of completeness, below we provide the most important relations used in this work:
\begin{eqnarray}
\hat{\Gamma}^{\epsilon}_{\mu\nu} &=& \Gamma^{\epsilon}_{\mu\nu} +C^{\epsilon}{}_{\mu\nu}, \\
\hat{R}_{\mu\nu}&=& R_{\mu\nu}- 2\nabla_{[\mu}C^{\gamma}_{\ \gamma]\nu}+ 2C^{\lambda}_{\ \nu[\mu}C^{\gamma}_{\ \gamma]\lambda},
\end{eqnarray}
where $\hat{\Gamma}^{\epsilon}_{\mu\nu}$ and $\Gamma^{\epsilon}_{\mu\nu}$ are Christoffel symbols and, $\hat{R}_{\mu\nu}$ and $R_{\mu\nu}$ are Ricci tensors of the metrics $\hat g_{\mu\nu}$ and $g_{\mu\nu}$, respectively. The auxiliary tensor $C^{\epsilon}_{\ \mu \nu}$ is given by
\begin{eqnarray}
C^{\epsilon}_{\ \mu \nu} &=& \frac{1}{\alpha}\Big[  \delta^{\epsilon}{}_{(\mu}\nabla_{\nu)}\alpha + \beta V^{\epsilon}\nabla_{(\mu}V_{\nu)} + V^{\epsilon} V_{(\mu}\nabla_{\nu)}\beta \nonumber\\[1ex]
&& +\beta V_{(\mu}\nabla_{\nu)}V^{\epsilon}  - \beta V_{(\mu}\nabla^{\epsilon}V_{\nu)} -\frac{1}{2} g_{\mu\nu}\nabla^{\epsilon}\alpha\nonumber\\[1ex]
&&-\frac{1}{2}V_{\mu}V_{\nu} \nabla^{\epsilon}\beta\Big]+ \frac{\beta\,V^{\epsilon}}{2\alpha^{2}}\Big[g_{\mu\nu} V^{\lambda}\nabla_{\lambda}\alpha - 2V_{(\mu}\nabla_{\nu)}\alpha\nonumber\\[1ex]
&&+ \beta V^{\lambda}\nabla_{\lambda}(V_{\mu}V_{\nu}) + V_{\mu}V_{\nu}V^{\lambda}\nabla_{\lambda}\beta\Big],
\end{eqnarray}
where $\nabla_{\mu}$ means covariant derivative and we denote $A_{[\mu\nu]}\equiv \frac{1}{2}(A_{\mu\nu} - A_{\nu\mu})$ and $A_{(\mu\nu)}\equiv \frac{1}{2}(A_{\mu\nu} + A_{\nu\mu})$.


\section{Singularity theorems and conformal transformation revisited}
Before approaching the case of disformal transformations, we would like to address a few comments about the singularity theorems for null congruence of curves and conformal transformations. Once they are particular cases of the disformal transformations, the results obtained here will also be true there. Furthermore, the discussion about the validity of the singularity theorems when different metrics are conformally related is sufficiently wide (see \cite{clarke,clarke_book,beem} for more details) such that it can shed some light to a possible extension in the context of the disformal transformations.

Let us start with a manifold $\mathcal{M}$ endowed with a metric $g_{\mu\nu}$ and consider an arbitrary positive function $\alpha \in C^{\infty}(\mathcal{M})$. Then, we consider another space-time $(\mathcal{M}, \hat{g})$ whose metric is
\begin{equation}
\label{conf_trans_met}
\hat g_{\mu\nu}=\alpha\, g_{\mu\nu}.
\end{equation}
It is straightforward to show that a null geodesic with tangent vector $k^{\mu}$ in $g_{\mu\nu}$ still satisfies the geodesic equation in the metric $\hat g_{\mu\nu}$. Thus, after the conformal transformation, we have
\begin{equation}
\label{geod1}
k^{\nu}\hat{\nabla}_{\nu}k^{\mu}=\frac{\dot{\alpha}}{\alpha}k^{\mu},
\end{equation}
where $\hat{\nabla}_{\mu}$ means covariant derivative with respect to $\hat g_{\mu\nu}$ and $\dot{\alpha}\doteq k^{\mu}\nabla_{\mu}\alpha$. Considering $\lambda$ as the affine parameter of the null geodesics with respect to $g_{\mu\nu}$, a redefinition of it like
\begin{equation}
\tilde{\lambda}(\lambda)=\int_0^{\lambda} \alpha(p)dp
\end{equation}
puts the geodesic equation in its standard form  $\hat{k}^{\mu}\widehat{\nabla}_{\mu}\hat{k}^{\nu}=0$ in the metric $\hat g_{\mu\nu}$,
where now the tangent vector is redefined as $\hat{k}^{\mu}=k^{\mu}/\alpha$. From this, it is worth to recall that causality (light-cones) is preserved by conformal transformations, which will not be true for disformal ones.

Notwithstanding, the first hypothesis of Penrose's singularity theorem for null congruences concerns the full projection of the Ricci tensor along the tangent vector of {\it any} null curve. So, by a direct computation of the Ricci tensor associated to $\hat g_{\mu\nu}$ doubly contracted with an arbitrary light-like vector $k^{\mu}$, we obtain
\begin{equation}
\label{conf-focus}
\widehat{R}_{\mu\nu}k^{\mu}k^{\nu}=R_{\mu\nu}k^{\mu}k^{\nu}+\frac{3}{2}\left(\frac{\dot{\alpha}}{\alpha}\right)^2-\frac{\ddot\alpha}{\alpha},
\end{equation}
or in terms of the auxiliary variable
$u=1/\sqrt{\alpha}$, it becomes
\begin{equation}
\label{conf-focus_u}
\widehat{R}_{\mu\nu}k^{\mu}k^{\nu}=R_{\mu\nu} k^{\mu} k^{\nu} + 2\,\frac{\ddot u}{u}.
\end{equation}
Note that the RHS of this equation is similar to the equation of a harmonic oscillator with time dependent frequency. In order to satisfy the {\it focusing condition}\footnote{From now on, we shall use this terminology to refer to the quantity $R_{\mu\nu} k^{\mu} k^{\nu}\geq0$ in order to avoid allusion to any theory of gravitation.}, i.e., $\widehat{R}_{\mu\nu}\hat k^{\mu}\hat k^{\nu}\geq0$, it is sufficient to require that $2\ddot u/u\geq - R_{\mu\nu}k^{\mu}k^{\nu}$ is valid for all null vector $k^{\mu}$. In particular, if $({\cal M},g)$ is Ricci flat, then $u(\lambda)$ must be an concave function of the affine parameter. Although the conformal transformation preserves the null geodesics of the space-time, note that the focusing condition gets altered. This allows us to map a singular space-time into a non-singular one through a suitable choice of the conformal function $\alpha$ (see further details in Ref.\ \cite{clarke}).

In order to verify the existence of closed trapped surfaces $\Sigma$, it can be done through the analysis of the sign of the norm of the mean curvature vector $H_{\mu}$ on $\Sigma$ (see appendix \ref{senov}). In fact, we need to check whether the following scalar function (defined by Eq.\ \ref{xi}) is positive
\begin{equation}
\hat{\xi}=-\hat{g}^{ab}\hat{H}_a\hat{H}_b,
\end{equation}
where $\hat{g}^{ab}$ are the components of the inverse conformal metric tensor. The lowercase Latin indices $(a,b,\ldots)$ denote the space-time coordinates running from $(0,1)$ while uppercase Latin indices $(A,B,\ldots)$ denote coordinates for $\Sigma$ running from $(2,3)$, and $\hat{H}_a$ are the conformal components of the mean curvature tensor calculated from the extrinsic curvature of the closed trapped surface.

From the definition of the scalar function $U=U(x)$ as $e^{U(x)}=\sqrt{{\rm det}\, \gamma}$, a straightforward calculation yields
\begin{equation}
\hat{H}_{\mu}=H_{\mu}+\frac{1}{\alpha}\delta^a_{\mu}(\alpha_{,a}-\gamma^{AB}\alpha_{,B}g_{aA}),
\end{equation}
where $\gamma_{AB}$ is the induced metric of the closed surface candidate as a trapped surface. Note that $\hat{H}_{\mu}$ can be written in terms of quantities defined in $g_{\mu\nu}$ and the conformal function $\alpha$. Therefore, the hypotheses {\bf (i)} and {\bf (iii)} of Penrose's singularity theorem applied for $\hat g_{\mu\nu}$ can be reformulated standing conditions over the space-time $(\mathcal{M},g)$ exclusively, as we state in the following

\begin{theorem}
Let $(\mathcal{M},g) \mapsto (\mathcal{M},\hat g)$ be a conformal transformation with $\hat g$ given by Eq.\ (\ref{conf_trans_met}), such that the space-time $(\mathcal{M},\hat g)$ admits a non-compact Cauchy surface. Let $R_{\mu\nu}$ denote the Ricci tensor of $(\mathcal{M},g)$ and $\lambda$ be the parameter along the light-like curves whose tangent vector is $k^{\mu}$.  If
\begin{enumerate}
    \item $2\sqrt{\alpha}\,\frac{d^2}{d\lambda^2}\left(\alpha\right)^{-\frac{1}{2}}\geq - R_{\mu\nu}k^{\mu}k^{\nu}$, for {\bf all} $k^{\mu}$;
    \item There exists a closed surface $\Sigma$ such that $\hat{\xi}>0$,
\end{enumerate}
then $(\mathcal{M},\hat g)$ is null geodesically incomplete.
\end{theorem}

It should be emphasized that this formulation of Penrose's singularity theorem as applied here for $(\mathcal{M},\hat g)$ allows one to test the first and third hypothesis of the theorem without knowledge of any geometrical property of such space-time.


\section{Singularity theorem and disformal transformation}
Recently, some of us have shown \cite{clb16} that a disformal transformation of the kind (\ref{cov}) can be seen as the action of an operator on vector fields over $\mathcal{M}$, i.e. $\overrightarrow{D}: \Gamma(T\mathcal{M})\,\rightarrow \, \Gamma(T\mathcal{M})$, such that its action can be split into two parts: the conformal and the purely disformal ones. Once the action of the conformal group on the hypothesis of the singularity theorems has been discussed previously, we can focus on the disformal component of the full transformation. Fortunately, such purely disformal transformation was widely studied in the literature under the terminology of Kerr-Schild transformation in the context of general relativity (some reviews on this topic can be found in \cite{stephani,Senovilla:2011fk,bini}).

In a coordinate system, we can represent a purely disformal transformation as
\begin{equation}
\widehat{g}_{\mu\nu}=g_{\mu\nu}+\epsilon d_{\mu}d_{\nu},
\label{fin_cov}
\end{equation}
whose inverse metric is
\begin{equation}
\widehat{g}^{\mu\nu}=g^{\mu\nu}-\epsilon d^{\mu}d^{\nu},
\label{fin_contrav}
\end{equation}
where $d^{\mu}$ denotes the light-like {\it disformal vector} with respect to both metrics and $\epsilon=\pm1$. While $\epsilon=+1$ provides the Kerr-Schild metrics as we find in the literature, the case $\epsilon=-1$ is also interesting because a given null vector $k^{\mu}$ with respect to the background metric $g_{\mu\nu}$ can be either time-like or light-like with respect to the disformal metric $\hat g_{\mu\nu}$, lying within the light-cone of the background metric. Whatever the case, the choice of $\epsilon$ can be done without loss of generality\footnote{For more details concerning the causality issue in disformal metrics see \cite{clb16}, paying attention to the different signature convention employed there.}. However, the class of vectors we must deal with are here those whose the norm is zero with respect to $\hat g_{\mu\nu}$, that is, we are interested in the light-like vectors $k^{\mu}$ tangent to the light-like curves in the space-time $(\mathcal{M}, \hat g)$:
\begin{equation}
\label{null_cond}
\hat g_{\mu\nu}k^{\mu}k^{\nu}=0\qquad \Longrightarrow \qquad g_{\mu\nu}k^{\mu}k^{\nu}=-\epsilon \phi^2,
\end{equation}
where $\phi\doteq d_{\mu}k^{\mu}$. Thus, for $\phi\neq0$, $k^{\mu}$ will be time-like if $\epsilon=+1$ or space-like if $\epsilon=-1$. For $\phi=0$, $k^{\mu}$ is light-like in both metrics.

Using the disformal metric (\ref{fin_cov}), the focusing term can be straightforwardly calculated and expressed in terms of the background metric, yielding
\begin{equation}
\label{foc_cond_dis_time}
\begin{array}{lcl}
\widehat{R}_{\mu\nu}k^{\mu} k^{\nu} &=& \left[R_{\mu\nu} + \frac{1}{2}d^{\,'}_{\mu}d^{\,'}_{\nu} +\epsilon D^{\alpha}{}_{\alpha}\,D_{\mu\nu} + \epsilon D'_{\mu\nu}\right.\\[2ex]
 &&\left.- 2\epsilon D_{[\nu\alpha]}D_{\mu}{}^{\alpha}\right]k^{\mu}k^{\nu} + 2\phi\,k^{\mu}\left(d^{\nu}D'_{[\mu\nu]} \right.\\[2ex]
 &&\left.+ D^{\nu}{}_{[\nu}d'{}_{\mu]}  -\epsilon \nabla^{\nu}D_{[\mu\nu]}\right)\\[2ex]
&&+ \frac{\phi^2}{2}\left(2 D^{\mu\nu}D_{[\mu\nu]} - \epsilon g^{\mu\nu}d^{\,'}_{\nu}d^{\,'}_{\mu}\right),
\end{array}
\end{equation}
where we use that $\epsilon^2=1$ and we define $D_{\mu\nu}\doteq\nabla_{\nu}d_{\mu}$, $d^{\,'}_{\mu}\doteq d^{\alpha}D_{\mu\alpha}$ and $D'_{\mu\nu}\doteq d^{\alpha}\nabla_{\alpha}D_{\mu\nu}$. In particular, if $\phi=0$ ($k^{\mu}$ is parallel to $d^{\mu}$) then Eq.\ (\ref{foc_cond_dis_time}) reduces a lot, but still the focusing condition may be satisfied in only one of the space-times, for instance, $R_{\mu\nu}k^{\mu}k^{\nu}\geq0$ in $g_{\mu\nu}$ cannot ensure that $\hat R_{\mu\nu}k^{\mu}k^{\nu}\geq0$ will be valid in $\hat g_{\mu\nu}$.

Now, we analyze the appearance of marginally closed trapped surfaces, which might indicate the existence of a trapped region in this space-time. Let us consider a two-dimensional compact hypersurface $\Sigma$ and a pair of null congruences $l^{\pm}_{\mu}$ with respect to $\hat g_{\mu\nu}$ orthogonal to $\Sigma$ satisfying $l^{\pm}_{\mu}\hat l^{\pm\mu}=0$ and $l^{\pm}_{\mu}\hat l^{\mp\mu}=-1$, where $\hat l^{\pm\mu}\doteq \hat g^{\mu\nu}l^{\pm}_{\nu}$.
The expansion coefficient $\hat\theta^{\pm}$ of these congruences can be written in terms of the corresponding expansion coefficient $\theta^{\pm}$ in the metric $g_{\mu\nu}$ as
\begin{equation}
\label{map_theta}
\hat\theta^{\pm}=\theta^{\pm} + \kappa^{\pm} + \epsilon d^\nu l^{\pm}_{\nu}\nabla_{\mu} d^{\mu} + \epsilon d^{\mu}\nabla_{\mu}(d^\nu l^{\pm}_{\nu}).
\end{equation}
Since $l^{\pm}_{\mu}$ is not necessarily mapped into an affinely parameterized curve, we introduce the parameter $\kappa^{\pm}$ to account for it. In order that $\Sigma$ be a closed trapped surface with respect to the disformal metric, the RHS of Eq.\ (\ref{map_theta}) should vanish. Then, identifying $d^{\mu}\nabla_{\mu}$ as the absolute derivative along the integral curve of $d^{\mu}$, we can solve Eq.\ (\ref{map_theta}) for $\psi^{\pm}\doteq d^\nu l^{\pm}_{\nu}$, as follows

\begin{equation}
\label{sol_map_theta_0}
\psi^{\pm}=\frac{C-\epsilon\int[\theta^{\pm} + \kappa^{\pm}](u)e^{\int_{u} \nabla_{\mu} d^{\mu}(\tilde u)d\tilde u}du}{e^{\int \nabla_{\mu} d^{\mu}(v)dv}},
\end{equation}
where $C$ is an integration constant. We emphasize that the integrals are calculated along the integral curves of $d^{\mu}$. Considering that the expansion factor $\theta$ is a scalar that describes the change in volume of a sphere of test particles centered on a given curve of the null congruence, the argument within the integral on the numerator of Eq.\ (\ref{sol_map_theta_0}) might be regarded as an overall {\it measure} change in the manifold when considering $\hat{g}$ instead of $g$. This is the sort of study performed in the realm of geometric analysis and geometric measure theory. For now, we abstain ourselves from delving into this problem.

Another way to study how the closed trapped surfaces are modified by a disformal transformation is again through the formalism presented in \ref{senov}. There, we only need to verify how the scalar (\ref{xi}) is altered by such transformation and try to solve a specific equation for it. Thus, we start by making a $2+2$ decomposition of the space-time associated to the disformal metric $\hat g_{\mu\nu}$ writing the squared line element in the following form
\begin{equation}
\begin{array}{lcl}
ds^2&=&(g_{ab}+\epsilon\, d_ad_b)dx^a dx^b+2(g_{aA} +\epsilon\, d_ad_A)dx^a dx^A \\[2ex]
&&+ (g_{AB}+\epsilon\, d_Ad_B) dx^A dx^B\, ,
\end{array}
\end{equation}
where $d_a$ and $d_A$ are, respectively, the $(0,1)$-components and the $(2,3)$-components of the disformal vector $d_{\mu}$ and the coordinates $\left\{x^A\right\}$ label the closed space-like surface $\Sigma$ candidate as a trapped surface. This decomposition allows us to identify the disformal components of the space-time metric as
\begin{equation}
\begin{array}{lcl}
\hat{g}_{ab}&=&g_{ab}+\epsilon\, d_ad_b,\\[1ex]
\hat{g}_{aA}&=&g_{aA}+\epsilon\, d_ad_A,\\[1ex]
\hat{g}_{AB}&=&g_{AB}+\epsilon\, d_Ad_B.
\end{array}
\end{equation}
With this decomposition, we can write the mean curvature covector using Eq.\ (\ref{mean_curv_vec}) and then we can construct a scalar given by Eq.\ (\ref{xi}) which indicates the formation of a closed trapped surface when it assumes positive values at some space-time region. Now, we describe the procedure to do so.

From a straightforward calculation, we first find the determinant of the disformal components of $\hat{\gamma}_{AB}$ as
\begin{equation}
\label{det_gamma_hat}
{\rm det}\, \hat{\gamma}=(1+\epsilon\, d^Ad_A){\rm det}\, \gamma.
\end{equation}
where we define the symbol $d^A\doteq \gamma^{AB}d_B$, with $\gamma_{AB}$ as the induced metric on $\Sigma$. Then, we define an auxiliary function $F(x)\doteq\sqrt{1+\epsilon\, d^Ad_A}$, such that the derivative of $U(x)$ in the disformal metric can be written down as
\begin{equation}
\label{h-u-function}
\hat{U}_{,a}=U_{,a}+\frac{F_{,a}}{F}.
\end{equation}
Now, we need to transform the term ${\rm div}\, {\bf g}_a$. Defining $\widehat{{\rm div}\, {\bf g}_a}\doteq(\sqrt{{\rm det}\, \hat{\gamma}}\, \hat{\gamma}^{AB}\, \hat{g}_{aA})_{,B}/\sqrt{{\rm det}\, \hat{\gamma}}$, a direct computation yields
\begin{equation}
\hat{\gamma}^{AB}\hat{g}_{aA}=\gamma^{AB}g_{aA}+\frac{\epsilon}{F^2}d^B\left(d_a-d^Ag_{aA}\right),
\end{equation}
and, thus, we find
\begin{equation}
\label{disf_div}
\widehat{{\rm div}\, {\bf g}_a}={\rm div}\, {\bf g}_a+\gamma^{AB}g_{aA}\frac{F_{,B}}{F}+\frac{\epsilon}{F}{\rm div}\, {\bf I}_a,
\end{equation}
where we have introduced the auxiliary covector ${\bf I}_a\doteq I_{aC}dx^C$, with
\begin{equation}
\label{i-function}
I_{aC}=\frac{d_C}{F}(d_a-d^Ag_{aA}).
\end{equation}
Finally, the mean curvature covector defined in the disformal metric is
\begin{equation}
\label{h_hat}
\hat{H}_{\mu}=\delta^a_{\mu}(\hat{U}_{,a}-\widehat{{\rm div}\, {\bf g}_a}),
\end{equation}
which allows us to compute its corresponding norm as being
\begin{equation}\label{h-xi-scalar}
\hat{\xi}=-\hat{g}^{ab}\hat{H}_a\hat{H}_b.
\end{equation}
Thus, a marginally trapped surface is formed when this scalar vanishes and, with the help of Eqs.\ (\ref{det_gamma_hat})-(\ref{h_hat}), this can be verified using solely the background metric and the disformal vector, without mention to the disformal metric. This means that the appearance of a closed trapped surface in this case would be due to the presence of a preferred direction provided by the disformal vector.

In summary, the restriction of Penrose's singularity theorem to disformal transformations that map a given space-time to another one preserving the causality condition leads to the following:
\begin{theorem}\label{theo2}
Let $(\mathcal{M},g) \mapsto (\mathcal{M},\hat g)$ be a disformal transformation given by Eq.\ (\ref{fin_cov}), such that the space-time $(\mathcal{M},\hat g)$ has a non-compact Cauchy surface. Let $R_{\mu\nu}$ denote the Ricci tensor of $(\mathcal{M},g)$.  If
\begin{enumerate}
    \item For all vector $k^{\mu}$ satisfying $g_{\mu\nu}k^{\mu}k^{\nu}=-\epsilon\,\phi^2$, we have
    \begin{equation}
    \begin{array}{l}
    R_{\mu\nu}k^{\mu}k^{\nu}\geq -\left( \frac{1}{2}d^{\,'}_{\mu}d^{\,'}_{\nu} +\epsilon D^{\alpha}{}_{\alpha}\,D_{\mu\nu} +\epsilon D'_{\mu\nu}\right.\\[2ex]
    \left.- 2\epsilon D_{[\nu\alpha]}D_{\mu}{}^{\alpha}\right)k^{\mu}k^{\nu}-2k^{\mu}\phi\left(d^{\nu}D'_{[\mu\nu]} + D^{\nu}{}_{[\nu}d'{}_{\mu]}\right.\\[2ex]
    \left. - \epsilon \nabla^{\nu}D_{[\mu\nu]}\right) -\frac{\phi^2}{2}\left(2 D^{\mu\nu}D_{[\mu\nu]} - \epsilon g^{\mu\nu}d^{\,'}_{\nu}d^{\,'}_{\mu}\right),
    \end{array}
    \end{equation}
where $\phi=d_{\mu}k^{\mu}$;
    \item There exists a closed surface $\Sigma$ in $(\mathcal{M},\hat g)$ such that $\hat{\xi}>0$,
\end{enumerate}
then $(\mathcal{M},\hat g)$ is null geodesically incomplete.
\end{theorem}

The extension of Theorem \ref{theo2} to the class of disformal metrics given by Eq.\ (\ref{cov}) can be achieved by making a conformal transformation of the Kerr-Schild metric (\ref{fin_cov}) with $\epsilon=1$ and a replacement of $d_{\mu}$ by $\sqrt{\frac{\beta}{\alpha}}\,d_{\mu}$ everywhere along this section.

\section{Applications to static and spherically symmetric space-times}\label{appl}
Now, we shall apply the previous results to spherically symmetric space-times that are disformally related, without assuming any theory of gravitation a priori. The idea for dealing with this family of space-times lies on the fact that the spherically symmetric space-times are widely studied in the context of gravitational collapse and black hole formation which are the most common issues where the Penrose singularity theorem is applied, besides it is very enlightening to work with as an example of the framework developed here. In this vein, one may have a better understanding about the features that one should expect from the dynamics of the metric in order to achieve a desired behaviour (singular or not) in a gravitational collapse scenario.

According to Theorem 2, we are capable to decide, for a Kerr-Schild metric like Eq.\ (\ref{fin_cov}) satisfying the causality condition, if a singularity can emerge through an operational test of the focusing condition and controlling the trapped surface formation. Again, it should be noticed that this can be done by using the tools defined strictly in the background geometry\footnote{Again, the only condition that concerns the disformal metric is the existence of a global Cauchy surface. Otherwise, the singularity can be avoided even if there are trapped surfaces and a Cauchy horizon with the null focusing condition being satisfied (see details in \cite{borde1,borde2}).}. Thus, the only assumptions will be that the background metric is the flat Minkowski space ($g_{\mu\nu}=\eta_{\mu\nu}$) and that the disformal vector preserves both the time-like Killing vector and the spherical symmetry.

For late convenience, we start with the Minkowski metric in spherical coordinates $(v,r,\theta,\varphi)$, where $v$ is a light-like coordinate. Then, we apply the disformal transformation to it, such that the line element with the disformal metric becomes
\begin{equation}
\label{dis_schw}
\widehat{ds^2}=[-1+f^2(r)]dv^2 + 2\, dv\, dr +r^2\sin^2 \theta\, d\varphi^2 + r^2\, d\theta^2.
\end{equation}
Note that the light-like disformal vector is given by $d^{\mu}=f(r)\delta^{\mu}_r$ while its corresponding covector is $d_{\mu}=f(r)\delta_{\mu}^v$, as required by the symmetries.

In this case, it is straightforward to show that $d^{\mu}$ satisfy the geodesic equation in $\eta_{\mu\nu}$ and its covariant derivative in this metric admits a simple matrix representation given by
\begin{equation}
[D^{\mu}{}_{\nu}]={\rm diag}\left(0,\frac{df}{dr},\frac{f}{r},\frac{f}{r}\right).
\end{equation}

In order to calculate the RHS of the focusing term expressed by Eq.\ (\ref{foc_cond_dis_time}), without entering into the details about the geometrical properties of the disformal metric (\ref{dis_schw}), we shall calculate the covariant derivative of $D^{\mu}{}_{\nu}$ with respect to the Minkowski metric, and then, project it along the disformal vector. This also has a simple matrix form as
\begin{equation}
[D'^{\mu}{}_{\nu}]=f(r)\,{\rm diag}\left(0,\frac{d^2f}{dr^2},\frac{r\frac{df}{dr}-f}{r^2},\frac{r\frac{df}{dr}-f}{r^2}\right).
\end{equation}
Finally, we need the divergence of $D^{\mu}{}_{\nu}$ with respect to its contravariant index, which is
\begin{equation}
\nabla_{\mu}D^{\mu}{}_{\nu}=\left(0,\frac{r^2\frac{d^2f}{dr^2} +2r\frac{df}{dr}-2f}{r^2},0,0\right).
\end{equation}
Recall that $\nabla_{\mu}$ is calculated according to the Minkowski metric. With these quantities, we can compute all terms involving the disformal vector and its covariant derivative in the RHS of Eq.\ (\ref{foc_cond_dis_time}), yielding
\begin{equation}
\label{focus_sph_sym}
\hat R_{\mu\nu}k^{\mu}k^{\nu}=\phi^2\left[f\frac{d^2f}{dr^2} + \left(\frac{df}{dr}\right)^2+\frac{2f}{r}\frac{df}{dr}\right],
\end{equation}
where we have used that the class of vectors $k^{\mu}$ satisfying Eq.\ (\ref{null_cond}) always admits adapted coordinates such that its angular components $k^{\theta}$ and $k^{\varphi}$ vanish in virtue of the spherical symmetry.

If we impose that $\hat R_{\mu\nu}k^{\mu}k^{\nu}=0$ (the lower bound for the focusing condition), then we get a second-order differential equation for $f(r)$, which can be solved, leading to
\begin{equation}
\label{sol_f}
f_0^{\pm}(r)=\pm\sqrt{C_0+\frac{C_1}{r}},
\end{equation}
where $C_0$ and $C_1$ are integration constants. It is curious that the family of functions given by $f_0^{\pm}(r)$, for each choice of $C_0$ and $C_1$, has integration constants with physical meaning: $C_0\neq0$ yields a class of asymptotically non-flat metrics with non-vanishing curvature tensor which has no correspondence in the realm of general relativity; while $C_1$ is related to the mass of the compact object source of the gravitational field.

\begin{figure*}
\begin{center}
\includegraphics[width=55mm,height=45mm]{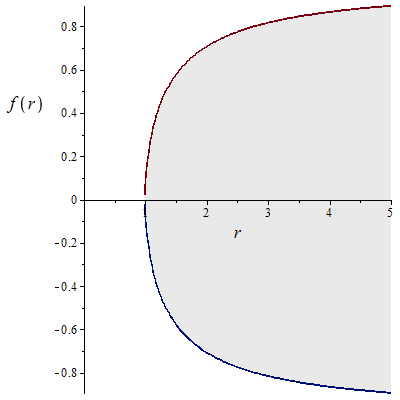}
\includegraphics[width=55mm,height=45mm]{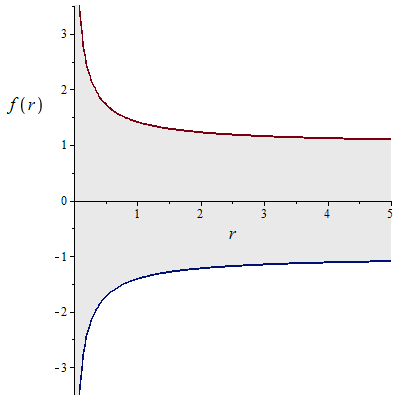}
\includegraphics[width=55mm,height=45mm]{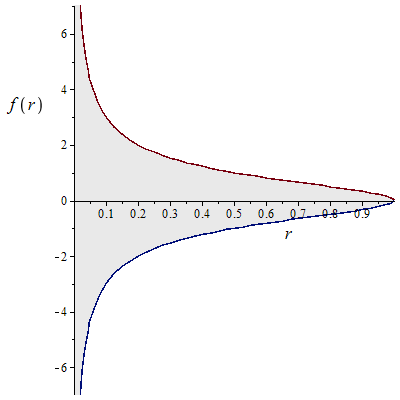}
\end{center}
\caption{Plots of $f_0^{\pm}(r)$. The shaded region indicates where the focusing condition is valid. On left, it is obtained for $C_0>0$ and $C_1<0$. On center, it was chosen $C_0,C_1>0$. On right, it was set $C_0<0$ and $C_1>0$. For both $C_0,C_1<0$, $f(r)$ is purely imaginary.}
\label{fig1}
\end{figure*}

If one takes small deviation $\delta^{\pm}$ of each branch of $f_{0}^{\pm}(r)$, for instance,  $f^{\pm}(r)= f^{\pm}_0(r)+\delta^{\pm}$, with $f^{\pm}_0(r)$ given by Eq.\ (\ref{sol_f}), then the focusing condition will be satisfied only for a certain combination of $f^{\pm}_0$ and the sign of $\delta^{\pm}$. This sets a range in the domain of the radial coordinate. In Fig.\ (\ref{fig1}), we depicted the behaviour of $f_0^{\pm}(r)$ for some illustrative values of the constants $C_0$ and $C_1$, shading the region where the focusing condition is satisfied.

In order to apply Senovilla's approach for spherically symmetric space-times, it is convenient the matrix representation of the Minkowski metric split into $2\times2$ blocks as
\begin{equation}
\label{m-matrix}
[\eta_{\mu\nu}]=\left(
\begin{array}{cc}
[\eta_{ab}] & \mathcal{O}_2\\
\mathcal{O}_2 & [\eta_{AB}]
\end{array}\right),
\end{equation}
where $\mathcal{O}_2$ is a $2\times2$ zero matrix and
\begin{equation}
[\eta_{ab}]=\left(
\begin{array}{cc}
-1 & 1 \\ 1 & 0
\end{array}\right),\quad
{\rm and}\quad
[\eta_{AB}]=[\gamma_{AB}]=\left(
\begin{array}{cc}
r^2 & 0 \\ 0 & r^2\sin^2\theta
\end{array}\right),
\end{equation}
whose the inverse matrix is
\begin{equation}
[\eta^{\mu\nu}]=\left(
\begin{array}{cccc}
0 & 1 &  0 & 0 \\
1 & 1 & 0 & 0 \\
0 & 0 & 1/r^2 & 0 \\
0 & 0 & 0 & 1/r^2\sin^2\theta
\end{array}\right).
\end{equation}
In this case, the derivative of the function $U(x)$, given by Eq.\ (\ref{u-function}), reads $U_{,a}=(2/r)\delta^r_a$, while $\eta_{aA}=0$,
implying that ${\rm div}\, {\bf g}_a=0$ and, therefore, the scalar (\ref{xi}) is given by
\begin{equation}
\xi=-\eta^{rr}U_{,r}U_{,r}=-\frac{4}{r^2}<0,
\end{equation}
which is non-positive. Therefore, there are no closed trapped surfaces in the Minkowski space-time, as already expected.

Now, let us check the validity of our expression for the disformal scalar $\hat{\xi}$ given by Eq.\ (\ref{h-xi-scalar}), with the help of Eqs.\ (\ref{h-u-function}) and (\ref{disf_div}). First, we notice that the disformal vector has $A$-components equal to zero, that is, $d_A=0$. This implies that the RHS of Eq.\ (\ref{disf_div}) is also zero, and then, $\widehat{{\rm div}\, {\bf g}_a}\equiv0$. Since $F=1$, the remaining term is the derivative of the function $\hat U$ that is written as $\hat{U}_{,a}=U_{,a}$. Therefore,
\begin{equation}
\hat{\xi}=\xi + (d^r U_{,r})^2,
\end{equation}
with $d^r=\eta^{\mu r}d_{\mu}=\eta^{v r}d_v=-f(r)$, which is explicitly given by
\begin{equation}
\hat{\xi}=-\frac{4[1-f^2(r)]}{r^2}.
\end{equation}
Note that $\hat{\xi}$ is non-negative only for $f^2(r)\geq1$ and the existence of such trapped region was possible only due to the disformal transformation of the scalar $\xi$.


\section{Concluding Remarks}\label{V}
The issue of the geodesic completeness under the conformal transformations has been debated in the literature since the appearance of the first singularity theorem and it has been demonstrated that an appropriate choice of the conformal function can map any strongly causal space-time into a null geodesic complete one \cite{clarke,beem}. When trying to extend these results to the realm of the disformal transformation, the situation becomes more complicated once one has more degrees of freedom to deal with.

Notwithstanding, we have shown that if one applies a disformal transformation to a non-singular space-time $(\mathcal{M},g)$ satisfying the focusing condition and without trapped surfaces, the presence of singularities in the resulting space-time $(\mathcal{M},\hat{g})$ can be verified only through the disformal transformation of the focusing condition and of Senovilla's scalar $\xi$, assuming that $(\mathcal{M},\hat{g})$ admits a non-compact Cauchy surface, as we argued before.

The implications of our results for alternative theories of gravity are many. In special, we have shown in the previous section that the focusing condition of a static and spherically symmetric disformal metric can be tested straightforwardly using Eq.\ (\ref{focus_sph_sym}) and the formation of closed trapped surface occurs if $f^2(r)-1$ admit real roots for positive $r$. These conditions made testable two out of the three hypotheses of the Penrose singularity theorem. If one try to extend the procedure presented here to axially symmetric space-times, this should be done carefully since metrics of this kind may not admit a Cauchy surface (as it is the case of the Kerr metric).

In conclusion, it is worth to mention that the violation of the focusing condition seems to be crucial for the avoidance of singularities, as far as we could see. In this vein, it suggests a classification of the space-time singularities according to the satisfaction (or not) of the hypotheses of the singularity theorems, but this demands a further investigation.

\begin{acknowledgements}
We would like to thank the participants of the {\it PMAT Seminar} for their valuable comments on a previous version of this manuscript. IPL and LS are financially supported by the Coordena\c{c}\~ao de Aperfei\c{c}oamento de Pessoal de N\'ivel Superior - Brasil (CAPES).
\end{acknowledgements}

\appendix

\section{Characterization of trapped surfaces}\label{senov}

In this section, we revisit the approach developed by Senovilla in \cite{Senovilla:2002ma,Senovilla:2011fk}, in order to define an equation that characterizes a (marginally) trapped surface in such a way that it could be easily modified when the space-time metric is affected by a disformal transformation.
\par
In a four dimensional manifold $\mathcal{M}$, a hypersurface $\Sigma$ can be represented by an embedding $\Phi:\, \Sigma \mapsto \mathcal{M}$ with parametric equations
\begin{equation}\label{embedding}
x^{\mu}=\Phi^{\mu}(\lambda),
\end{equation}
where $\{x^{\mu}\}$ are local coordinates in $\mathcal{M}$ ($\mu=0,1,2,3$), and $\lambda$ represents the set of local coordinates $\{\lambda^A\}$ for $\Sigma$ ($A=2,3$).
\par
We can find the metric on $\Sigma$ by restricting the line element to displacements confined to the hypersurface, i.e., by finding the first fundamental form of the surface $\Sigma$ induced by the geometry of the manifold $\mathcal{M}$. In fact, the vectors
\begin{equation}
\vec{e}_A=\frac{\partial \Phi^{\mu}}{\partial \lambda^A}\partial_{\mu},
\end{equation}
are tangent to curves contained in $\Sigma$. If we define $e^{\mu}_{\ A}=\partial \Phi^{\mu}/\partial \lambda^A$, the first fundamental form of $\Sigma$ in $(\mathcal{M},g)$ is simply the pull-back of $g$ given by $\gamma=\Phi^* g$, which in coordinates $\{\lambda^A\}$ is
\begin{equation}
\gamma_{AB}=g_{\mu\nu}e^{\mu}_{\ A}e^{\nu}_{\ B}.
\end{equation}
Using Eq.(\ref{embedding}), the line element of the surface reads
\begin{equation}
\begin{array}{lcl}
ds^2_{\Sigma}&=&g_{\mu\nu}dx^{\mu}dx^{\nu}|_{\Sigma}=g_{\mu\nu}\frac{\partial\Phi^{\mu}}{\partial \lambda^A}\frac{\partial\Phi^{\nu}}{\partial \lambda^B}d\lambda^Ad\lambda^B\\[2ex]
&=&\gamma_{AB}d\lambda^A d\lambda^B.
\end{array}
\end{equation}
It is always possible to decompose the tangent space at a point $x\in \mathcal{M}$ as $T_x \mathcal{M}=T_x \Sigma\oplus T_x \Sigma^{\perp}$, such that
\begin{equation}\label{2-fundam}
\nabla_{\vec{e}_A} \vec{e}_B=\tilde{\Gamma}^C_{AB}\vec{e}_C-\vec{K}_{AB},
\end{equation}
where $\nabla$ is the Riemannian connection of the manifold $\mathcal{M}$, $\tilde{\Gamma}^C_{AB}$ are the Christoffel symbols associated with the induced metric on $\Sigma$, i.e., $\tilde{\nabla}_C\gamma_{AB}=0$, and $\vec{K}_{AB}$ is called shape tensor or second fundamental form vector of $\Sigma$ in $\mathcal{M}$. In fact, we can project the covariant derivative of a 1-form field ${\bf v}$ onto the vectors $\vec{e}_A$, which gives
\begin{equation}
e^{\mu}_{\ B}e^{\nu}_{\ A}\nabla_{\nu}v_{\mu}=\tilde{\nabla}_A\tilde{v}_B+v_{\mu}K^{\mu}_{AB},
\end{equation}
where $\tilde{v}_A\doteq v_{\mu}e^{\mu}_{\ A}$. The usual second fundamental form relative to a 1-form ${\bf n}$ normal to the surface $\Sigma$ is simply
\begin{equation}
\vec{K}_{AB}[{\bf n}]=n_{\mu}K^{\mu}_{AB}.
\end{equation}

Since $\Sigma$ is a two-dimensional, compact, space-like surface in a four-dimensional manifold, we can always find two vectors that are linearly independent and choose them to be future-directed and light-like everywhere on $\Sigma$. These vectors will characterize the trapped surface, once the region that is confined by this surface has the property of confining light rays (and also massive particles).

Let us denote these null vectors as $\vec{k}^{\pm}$, satisfying $k^+_{\ \mu}k^{-\mu}=-1$ and $k^{\pm}_{\ \mu}e^{\mu}_{\ A}=0$. With these vectors, we can decompose the shape tensor as
\begin{equation}\label{dec1}
\vec{K}_{AB}=-\left(K^{\mu}_{AB}k^-_{\ \mu}\right)\vec{k}^+ - \left(K^{\mu}_{AB}k^+_{\ \mu}\right)\vec{k}^-\, ,
\end{equation}
and define the {\it mean curvature vector} of $\Sigma$ as the trace of the shape tensor using the induced metric:
\begin{equation}
\vec{H}=\gamma^{AB}\vec{K}_{AB}.
\end{equation}

Since $\vec{H}$ still carries the index of the shape tensor, it is orthogonal to $\Sigma$, i.e., $H_{\mu}e^{\mu}_{\ A}\equiv 0$. The decomposition (\ref{dec1}) allows us to define the expansion coefficients of the future-directed light-like vectors through the mean curvature vector by
\begin{equation}
\label{vec-h}
\vec{H}=-\theta^- \vec{k}^+ -\theta^+ \vec{k}^-,
\end{equation}
where
\begin{equation}
\label{theta}
\theta^{\pm}\doteq\gamma^{AB}K_{AB}^{\mu} k^{\pm}_{\ \mu}.
\end{equation}
The mean curvature vector is of fundamental importance for our purposes, since {\it the sign of its norm will furnish a necessary and sufficient condition for $\Sigma$ to be a trapped surface}. So, now we shall focus in expressing $H_{\mu}H^{\mu}$ in a useful way for our disformal analysis.

Thus, let us assume without loss of generality that our space-like surface $\Sigma$ is described by the conditions $x^a=constant$, where $a=0,1$. Locally, the squared line element can be written as
\begin{equation}
ds^2=g_{ab}dx^a dx^b+2g_{aA} dx^a dx^A + g_{AB} dx^A dx^B\, ,
\end{equation}
where ${\rm det}\, g_{AB} > 0$. In this case, the embedding $\Phi$ is $x^a=\Phi^a=X^a=const.$ and $x^A=\Phi^A=\lambda^A$. From these definitions, we see that the first fundamental form of $\Sigma$ is simply $\gamma_{AB}=g_{AB}(X,\lambda)$ and the future-directed null 1-forms ${\bf k}^{\pm}$ become ${\bf k}^{\pm}=k^{\pm}_{\ b}\, dx^b|_{\Sigma}$, which only have indices $(0,1)$ due to the space-time decomposition.

Consider an auxiliary scalar function defined in terms of the determinant of the first fundamental form as
\begin{equation}
\label{u-function}
e^{U(x)}\doteq\sqrt{{\rm det}\gamma}.
\end{equation}
and define the 1-form
\begin{equation}\label{ga-function}
{\bf g}_a=g_{aA}dx^A.
\end{equation}
From Eq.\ (\ref{2-fundam}) we can describe the shape tensor from derivatives of the metric, deriving then from Eq.(\ref{theta}) an expression for $\theta^{\pm}$ as follows (see \cite{Senovilla:2002ma} for more details)
\begin{equation}
\theta^{\pm}=k^{\pm\, a}\left[U_{,a}-e^{-U}(e^{U}\, \gamma^{AB}g_{aA})_{,B}\right].
\end{equation}
From the identity
\begin{equation}\label{hat_div}
{\rm div}\, {\bf g}_a=\gamma^{AB}\tilde{\nabla}_B g_{aA}=\frac{1}{{\sqrt{{\rm det}\, \gamma}}}(\sqrt{{\rm det}\, \gamma}\, \gamma^{AB}\, g_{aA})_{,B},
\end{equation}
we express the mean curvature vector as
\begin{equation}
\label{mean_curv_vec}
H_{\mu}=\delta^a_{\mu}(U_{,a}-{\rm div}\, {\bf g}_a).
\end{equation}
Finally, we have that $\Sigma$ is a trapped surface if and only if
\begin{equation}
\label{xi}
\xi = -g^{bc}H_bH_c|_{\Sigma}
\end{equation}
is positive. A necessary condition for it to be a marginally trapped surface is that $\xi$ vanishes. This expression is very useful for our purposes, since it just depends on the metric tensor, which is the geometrical object that if affected by a disformal transformation.


\begin{thebibliography}{50}
\bibitem{belinski}
V. Belinski, \textit{On the cosmological singularity}, { Int.\ J.\ Mod.\ Phys.\ D} {\bf 23} 1430016 (2014).
\bibitem{haw_ellis}
S. W. Hawking and G. F. R. Ellis, \textit{The Large Scale Structure of Space-Time}. (Cambridge University Press, Cambridge, 1973).
\bibitem{penrose}
R. Penrose, \textit{Gravitational collapse and space-time singularities}, {Phys.\ Rev.\ Lett.} {\bf 14} 57 (1965).
\bibitem{Senovilla:2018aav}
  J.~M.~M.~Senovilla, \textit{Singularity Theorems and Their Consequences}, {Gen.\ Rel.\ Grav.}\  {\bf 30} 701 (1998).
\bibitem{martin}
M. Reiris and J. Peraza, \textit{A complete classification of S1-symmetric static vacuum black holes}, {Class.\ Quantum Grav.} {\bf 36} 225012 (2019).
\bibitem{magueijo04}
J. Magueijo and L. Smolin, \textit{Gravity's rainbow}, {Class.\ Quantum Grav.} {\bf 21} 1725 (2004).
\bibitem{clb16}
G. G. Carvalho, I. P. Lobo and E. Bittencourt, ``Extended disformal approach in the scenario of rainbow gravity'', {Phys.\ Rev.\ D} {\bf 93} 044005 (2016).
\bibitem{amelino01}
G. Amelino-Camelia, \textit{Relativity in space-times with short-distance structure governed by an observer-independent (planckian) length scale}, {Int. J. Mod. Phys. D} {\bf 11} 35 (2002).
\bibitem{beken_mond}
J. D. Bekenstein, \textit{Relativistic gravitation theory for the modified Newtonian dynamics paradigm}, {Phys.\ Rev.\ D} {\bf 70} 083509 (2004), [Erratum-ibid. D {\bf 71} 069901 (2005)].
\bibitem{scalartheory}
M. Novello, E. Bittencourt, U. Moschella, E. Goulart, J. M. Salim and J. D. Toniato, \textit{Geometric scalar theory of gravity}, {J.\ Cosm.\ Astro.\ Phys.} JCAP06 (2013) 014.
\bibitem{mota}
T. S. Koivisto, D. F. Mota and M. Zumalacarregui, \textit{Screening Modifications of Gravity Through Disformally Coupled Fields}, {Phys.\ Rev.\ Lett.} {\bf 109} 241102 (2012).
\bibitem{ip}
H. Y. Ip, J. Sakstein, F. Schmidt, \textit{Solar system constraints on disformal gravity theories}, {J.\ Cosm.\ Astro.\ Phys.}, {\bf 10} 051 (2015).
\bibitem{sak1}
J. Sakstein, \textit{Disformal theories of gravity: from the solar system to cosmology}, {J.\ Cosm.\ Astro.\ Phys.}, {\bf 12} 012 (2014).
\bibitem{sak2}
J. Sakstein, S. Verner, \textit{Disformal gravity theories: A Jordan frame analysis}, {Phys.\ Rev.\ D} {\bf 92}, no. 12, 123005 (2015).
\bibitem{rua}
N. Deruelle and J. Rua, \textit{Disformal transformations, veiled General Relativity and Mimetic Gravity}, {J.\ Cosm.\ Astro.\ Phys.}, {\bf 09} 002 (2014).
\bibitem{matarrese}
F. Arroja, N. Bartolo, P. Karmakar and S. Matarrese, \textit{The two faces of mimetic Horndeski gravity: disformal transformations and Lagrange multiplier}, {J.\ Cosm.\ Astro.\ Phys.}, {\bf 09} 051 (2015).
\bibitem{sunny1}
R. Myrzakulov, L. Sebastiani, S. Vagnozzi and S. Zerbini, \textit{Remarks on and cosmological extensions of covariant renormalizable gravity}, {Fund.\ J.\ Mod.\ Phys.} {\bf 8} 119 (2015).
\bibitem{sunny2}
R. Myrzakulov, L. Sebastiani, S. Vagnozzi and S. Zerbini, \textit{Static spherically symmetric solutions in mimetic gravity: rotation curves \& wormholes}, arXiv:[gr-qc]1510.02284.
\bibitem{dario}
D. Bettoni and S. Liberati, \textit{Disformal invariance of second order scalar-tensor theories: Framing the Horndeski action}, {Phys.\ Rev.\ D} {\bf 88} 084020 (2013).
\bibitem{miguel1}
M. Zumalac\'arregui and J. Garc\'ia-Bellido, \textit{Transforming gravity: From derivative couplings to matter to second-order scalar-tensor theories beyond the Horndeski Lagrangian}, {Phys.\ Rev.\ D} {\bf 89} 064046 (2014).
\bibitem{vernizzi}
J. Gleyzes, D. Langlois, F. Piazza and F. Vernizzi, \textit{Exploring gravitational theories beyond Horndeski}, {J.\ Cosm.\ Astro.\ Phys.}, {\bf 02} 018 (2015).
\bibitem{nov_bit_gordon}
M. Novello and E. Bittencourt, \textit{Gordon metric revisited}, {Phys.\ Rev.\ D} {\bf 86} 124024 (2012).
\bibitem{nov_bit_drag}
M. Novello and E. Bittencourt, \textit{Dragged metrics}, {Gen.\ Rel.\ Grav.} {\bf 45} 1005 (2013).
\bibitem{uzan}
S. Mukoyama, J.-P. Uzan, \textit{Emergence of the Lorentzian structure in classical field theory}, {Int.\ J.\ Mod.\ Phys.\ D} {\bf 22} 1342018 (2013).
\bibitem{yuan}
F. -F. Yuan, P. Huang, \textit{Induced geometry from disformal transformation}, {Phys.\ Lett.\ B} {\bf 744} 120 (2015).
\bibitem{brax1}
P. Brax and C. Burrage, \textit{Constraining disformally coupled scalar fields}, {Phys.\ Rev.\ D} {\bf 90} 104009 (2014).
\bibitem{brax2}
P. Brax, C. Burrage and C. Englert, \textit{Disformal dark energy at colliders}, {Phys.\ Rev.\ D} {\bf 92} 044036 (2015).
\bibitem{bitt_nov_faci}
E. Bittencourt S. Faci and M. Novello, \textit{Chiral symmetry breaking as a geometrical process}, {Int.\ J.\ Mod.\ Phys.\ A} {\bf 29} 1450145 (2014);
\bibitem{nov_bit}
M. Novello and E. Bittencourt, \textit{A proposal for the origin of the anomalous magnetic moment}, {Int.\ J.\ Mod.\ Phys.\ A} {\bf 29} 1450075 (2014);
\bibitem{erico12}
F. T. Falciano and E. Goulart, \textit{A new symmetry of the relativistic wave equation}, {Class.\ Quantum Grav.} {\bf 29} 085011 (2012).
\bibitem{erico13}
E. Goulart and F. T. Falciano, \textit{Disformal invariance of Maxwell's field equations}, {Class.\ Quantum Grav.} {\bf 30} 155020 (2013).
\bibitem{bitt15}
E. Bittencourt, I. P. Lobo and G. G. Carvalho, \textit{On the disformal invariance of the Dirac equation}, {Class. Quantum Grav.} {\bf32} 185016(2015).
\bibitem{poisson}
E. Poisson, \textit{A Relativist's Toolkit}, Cambridge University Press, Cambridge, UK (2004).
\bibitem{stephani}
H. Stephani, D. Kramer, M. A. H. MacCallum, C. Hoenselaers and E. Herlt, {\em Exact Solutions of Einstein's Field Equations}, 2nd ed. (Cambridge University Press, Cambridge, 2003).
\bibitem{lc19}
I. P. Lobo, G. G. Carvalho, \textit{The geometry of null-like disformal transformations}, {Int.\ J.\ Geom.\ Meth.\ Mod.\ Phys.} {\bf 16} 1950180 (2019).
\bibitem{clarke}
C. J. S. Clarke, \textit{On the geodesic completeness of causal space-times}, {\em Mathematical Proceedings of the Cambridge Philosophical Society} {\bf 69(2)} 319 (1971).
\bibitem{clarke_book}
C. J. S. Clarke, \textit{The Analysis of Space-Time Singularities}, Cambridge University Press, Cambridge, (1993).
\bibitem{beem}
J. K. Beem, \textit{Conformal changes and geodesic completeness}, {Commun.\ Math.\ Phys.} {\bf 49} 179 (1976).
\bibitem{bini}
D. Bini, A. Geralico and R. P. Kerr, \textit{The Kerr-Schild ansatz revisited}, {Int.\ J.\ Geom.\ Meth.\ Mod.\ Phys.}  {\bf 7} 693 (2010).
\bibitem{Senovilla:2011fk}
J. M. M. Senovilla,  \textit{Trapped surfaces},  {Int.\ J.\ Mod.\ Phys.\ D} {\bf 20} 2139 (2011)  arXiv:1107.1344 [gr-qc].
\bibitem{Senovilla:2002ma}
J. M. M. Senovilla, \textit{Trapped surfaces, horizons and exact solutions in higher dimensions}, {Class.\ Quantum Grav.} {\bf} 19 (2002) L113 arXiv:hep-th/0204005.
\bibitem{borde1}
A. Borde, \textit{Open and closed universes, initial singularities, and inflation}, {Phys.\ Rev.\ D} {\bf 50} 3692 (1994) .
\bibitem{borde2}
A. Borde, \textit{Regular black holes and topology change}, {Phys.\ Rev.\ D} {\bf 55} 7615 (1997).
\end{thebibliography}
\end{document}